\documentclass[aps,amssymb,article,pra,twocolumn,showpacs,showkeys,floatfix]{revtex4}

\usepackage{epsf}
\usepackage{amsmath}
\usepackage{graphicx}
\begin{document}
         
\title{Anomalous electron states}

\author{Boris I. Ivlev}

\affiliation{Instituto de F\'{\i}sica, Universidad Aut\'onoma de San Luis Potos\'{\i},\\ 
San Luis Potos\'{\i}, 78000 Mexico}

\begin{abstract}

By the certain macroscopic perturbations in condensed matter anomalous electron wells can be formed due to a local reduction of electromagnetic zero point energy. These wells are narrow, of the 
width $\sim 10^{-11}cm$, and with the depth $\sim 1MeV$. Such anomalous states, from the formal standpoint of quantum mechanics, correspond to a singular solution of a wave equation produced by 
the non-physical $\delta(\vec R)$ source. The resolution, on the level of the Standard Model, of the tiny region around the formal singularity shows that the state is physical. The creation of 
those states in an atomic system is of the formal probability $\exp(-1000)$. The probability becomes not small under a perturbation which rapidly varies in space, on the scale $10^{-11}cm$. In 
condensed matter such perturbation may relate to acoustic shock waves. In this process the short scale is the length of the standing de Broglie wave of a reflected lattice atom. Under electron 
transitions in the anomalous well (anomalous atom) $keV$ X-rays are expected to be emitted. A macroscopic amount of anomalous atoms, of the size $10^{-11}cm$ each, can be formed in a solid 
resulting in {\it collapsed matter} with $10^{9}$ times enhanced density.

\end{abstract} \vskip 1.0cm



\maketitle

\section{INTRODUCTION}
\label{intr}
Discrete energy levels of the electron in a potential well are shifted due to the interaction with photons (Lamb shift) \cite{LANDAU3}. This phenomenon can be interpreted through electron 
``vibrations'' with the mean displacement $\langle\vec u\rangle=0$ and the non-zero mean squared displacement $\langle u^2\rangle=r^{2}_{L}$, where the Lamb radius $r_L\simeq 10^{-11}cm$ 
\cite{WEL,MIG,KOL}. This is the electron fluctuation spreading in addition to the quantum mechanical uncertainty. Usually $r_L$ is much smaller than that uncertainty. This is the reason why the 
Lamb shift is relatively small. 

One can try to analyze the case when the quantum mechanical uncertainty, for some reasons, is smaller than $r_T$. This situation may occur when the wave function of a bare electron (in formal 
absence of the electron-photon interaction) is singular at some point $\psi\sim 1/R$. Such a solution is allowed by wave equations at $R\neq 0$. Then ``switching on'' the interaction with photons 
may result in two consequences. 

First, the bare electron mass will be renormalized converting into the physical one acquiring the correction \cite{FEYN2,LANDAU3}
\begin{equation}
\label{0}
\Delta m=m\,\frac{3e^2}{2\pi \hbar c}\ln\frac{\hbar}{lmc}\,, 
\end{equation}
where $l$ is the small cutoff distance. According to ideas of quantum gravity (see the review paper \cite{HOS}, references therein, and \cite{LEM}), the cutoff $l$ has no pure mathematical meaning 
but $l\sim \sqrt{\hbar G_g/c^3}\simeq 1.62\times 10^{-33}cm$. This is the Planck length which is the fundamental minimal length scale. Here $G_g$ is the gravitational constant. From this angle, the
mass correction (\ref{0}) is less than $20\%$.

Second, the singularity of the electron distribution will be smeared within the region $r_L$. 

That scenario is not realized in quantum electrodynamics. The kinetic energy terms $-(\hbar^2/2m)\nabla^2(1/R)$ for the bare electron is also singular as $\delta(\vec R)$. To support this solution
the singular point like source should be in the wave equation for bare electron. This additional source is not physical.

However, one can try to resolve the short distance scale where the point source $\delta(\vec R)$ is supposed to locate. Search of short scales leads to the mechanism of electron mass generation. 
As known, in the Standard Model electron mass 
\begin{equation}
\label{01}
m=\frac{Gv}{c^2}
\end{equation}
is determined (through the Yukawa coupling $G$) by the mean value $v$ of the Higgs field \cite{GLA,ENG,HIG,GUR,WEI,VOL}. Usually $v$ weakly depends on electron distribution. Let us formally 
consider the bare electron (with no weak bosons $W^{\pm}, Z$, photons, and a fluctuating part of the Higgs field). In this case the mean value the Higgs field $v$ can be disturbed on short distances 
by the above singular electron distribution. In turn, the singular part of $v$ (according to (\ref{01})) results in a singular bare electron mass which serves as a natural singularity source (instead
of the artificial $\delta(\vec R)$) in the wave equation for the bare electron (Sec.~\ref{sing}).

The subsequent inclusion of the fluctuating fields results, as in quantum electrodynamics, in the renormalization of the electron bare mass. In the Standard Model, besides the photon term in 
(\ref{0}), there are analogous ones due to the interaction with $W^{\pm}$ and $Z$ \cite{LEM}. As in quantum electrodynamics, the difference between bare and physical masses is small. This mass 
renormalization can be interpreted as renormalization of the Yukawa coupling $G$ \cite{LEM}. 

In addition to the usual renormalization of the Yukawa coupling $G$, there is the novel aspect of the problem. The resulting state is a superposition of ones with singularity positions shifted by
the vector $\vec u$ determined by fluctuating fields. Therefore the physical electron density includes (besides renormalization effects) the average $\langle n(\vec R-\vec u)\rangle$ with respect 
to all fluctuating positions $\vec u$. In the usual case this would correspond to the Lamb effect. Sweeping of $\vec u$, at a fixed $R$, provides a contribution also from short distances, where the 
Standard Model is not valid. However, there is the minimal length scale $l$, mentioned above, which serves as cutoff. For this reason, the electron density, obtained in that way, is smooth and
physical (Sec.~\ref{smoo}). 

Within the Standard Model singularity positions $\vec u$ are determined by fluctuations of weak bosons $W^{\pm}, Z$, the Higgs field, and photons. Only photons remain massless providing the main
contribution to the fluctuating $\vec u$. The related fluctuation radius is $r_L$ which is of the electron-photon origin. The small $r_L$ is proportional to $e^2/\hbar c$ as it should be. But the 
initial electron distribution is singular and therefore smearing of this distribution is a non-perturbative phenomenon on $e^2/\hbar c$. 

The resulting anomalous electron state originates from the singular one which is smeared out mainly due to the electron-photon interaction. That state is localized within the region 
$r_L\sim 10^{-11}cm$. According to uncertainty principle, this relates to the increase of the electron energy by $\hbar c/r_L\sim 1MeV$. That energy enhancement is compensated by the local (within 
$r_L$ radius) reduction of zero point energy of photons. This is equivalent to the certain well of the $MeV$ depth recalling formation of a well of the similar origin in the Casimir effect 
\cite{CAS,LANDAU3} (Sec.~\ref{smooA}). 

For the free electron (which is not restricted by some macroscopic potential) $r_L=\infty$. Therefore anomalous state does not exist in vacuum. In this case there is the usual Lehmann 
representation of the electron propagator according to quantum electrodynamics \cite{LANDAU3}. Coulomb attraction field of lattice sites in a solid may play role of restriction potential. 

As shown in this paper, anomalous states can be formed by usual macroscopic processes in solids, for example, by a propagation of acoustic shock waves. In this case the standing de Broglie wave of 
a lattice site produces the charge density with the spatial scale of $\sim r_L$. The related matrix element between the usual electron state in a crystal lattice and anomalous one becomes not 
exponentially small. 

It is unusual that by a macroscopic perturbation in condensed matter $MeV$ energy electron well may be formed due to the local reduction of electromagnetic zero point energy. Under electron 
transitions in the anomalous well the emission of $keV$ X-rays is expected. An emission of higher energy quanta, in the $MeV$ region, is principally possible and requires more studies.

The anomalous well with electrons on its energy levels is anomalous atom which is three orders of magnitude smaller than a usual one. If in a part of a solid all atoms undergo a transition to the 
anomalous state that macroscopic region increases its density $10^{9}$ times. This {\it collapsed matter} looks as a dramatically different concept. 
\section{GENERATION OF ELECTRON MASS}
\label{gen}
In the Standard Model masses of electron, other leptons, $W^{\pm}$ and $Z$ weak bosons, and quarks are generated by Higgs mechanism which involves the scalar Higgs field \cite{GLA,ENG,HIG,GUR,WEI}. 
Electron, as a fermion, acquires its mass by the connection between the fermion field $\psi$ and the Higgs field $\phi$. The Lagrangian 
\begin{equation}
\label{126}
L=i\hbar c\bar{\psi}\gamma^{\mu}\tilde{D}_{\mu}\psi-G\bar{\psi}\phi\psi+L_{H}(\phi)+L_g
\end{equation}
contains the Higgs part
\begin{equation}
\label{126a}
L_{H}(\phi)=\frac{1}{\hbar c}(D_{\mu}\phi)^+D^{\mu}\phi+\frac{1}{(\hbar c)^3}\left[\mu^2c^4\phi^+\phi-\lambda(\phi^+\phi)^2\right]
\end{equation}
and the gauge part $L_g$ that, for pure electromagnetic field, would be $-F^{\mu\nu}F_{\mu\nu}/16\pi$ where $F_{\mu\nu}=\partial_{\mu}A_{\nu}-\partial_{\nu}A_{\mu}$. The Yukawa term, depending on 
the coupling $G$, is written in (\ref{126}) in a schematic form. The covariant derivatives $\tilde{D}_{\mu}$ and $D_{\mu}$ contain, in addition to partial derivatives $\partial_{\mu}$, the parts
depending on gauge fields $W^{\pm}_{\mu}$, $Z_{\mu}$, and $A_{\mu}$. In (\ref{126}) $\gamma^{\mu}$ are the Dirac matrices. 

The isospinor $\phi=(0,v+h)$, besides the expectation value $v$, contains the fluctuation part $h$ with zero expectation value. The electron mass $m_0=Gv_0/c^2$ appears (in the Yukawa term) due to 
the finite expectation value $v_0=\mu c^2$ that relates to the ground state of $L_H$ \cite{GLA,ENG,HIG,GUR,WEI}. So the parameter $G=m/\mu$, where $\mu\sim 100GeV/c^2$, is the mass of the Higgs 
boson. One can estimate $G\sim 10^{-5}$. We normalize the Higgs field to have $\lambda=1/2$.

Instead of solving the whole problem with fluctuating fields of gauge bosons $W^{\pm}_{\mu}$, $Z_{\mu}$, $A_{\mu}$, and $h$ one can separate the problem by two steps. At the first step, the 
fluctuating fields are formally ``switched off''. Without gauge bosons in the isospinor doublet $(\nu_e,e)$ there is no neutrino-electron mixing. Therefore, at the first step, one can consider the 
electron bispinor only which is denoted below as $\psi$. At the second step, the fluctuating fields $W^{\pm}_{\mu}$, $Z_{\mu}$, $A_{\mu}$, and $h$ are to be included.

We start the first step with the equation 
\begin{equation}
\label{127}
\nabla^2v+\frac{1}{\hbar^2c^2}(\mu^2c^4v-v^3)=\frac{\hbar c}{2}G\bar{\psi}\psi
\end{equation}
for the expectation value $v$ of the Higgs field which follows from the mean field analogue of Eq.~(\ref{126a}). Here the right-hand side can be calculated according to Dirac quantum mechanics. In
Eq.~(\ref{01}) the electron mass $m=m_0+\delta m(\vec R)$ is variable in space $\vec R=\{\vec r,z\}$ according to variations of $v$. 

The electron spinors $\varphi$ and $\chi$, which form the total bispinor $\psi=(\varphi,\chi)$, satisfy the equations \cite{LANDAU3}
\begin{eqnarray}
\nonumber
&&\left[\varepsilon-U(\vec R)+i\hbar c\vec\sigma\nabla\right]\varphi=mc^2\chi\\
\label{128}
&&\left[\varepsilon-U(\vec R)-i\hbar c\vec\sigma\nabla\right]\chi=mc^2\varphi.
\end{eqnarray}
Here $\varepsilon$ is the total relativistic energy and $\vec\sigma$ are Pauli matrices. In (\ref{128}) fluctuation electromagnetic field is ``switched off''. 

It follows from Eq.~(\ref{128}) that
\begin{equation}
\label{129}
\Theta=-\frac{i\hbar c\vec{\sigma}\nabla\Phi}{\varepsilon-U+mc^2}\,,
\end{equation}
where $\Phi=(\varphi+\chi)/\sqrt{2}$ and $\Theta=(\varphi-\chi)/\sqrt{2}$.
The spinor $\Phi$ satisfies the equation
\begin{equation}
\label{130}
-\nabla^2\Phi+\frac{\nabla\beta}{1+\beta}\left(\nabla\Phi-i\vec\sigma\times\nabla\Phi\right)+\frac{m^2c^2}{\hbar^2}\Phi=\frac{(\varepsilon-U)^2\Phi}{\hbar^2c^2}\,,
\end{equation}
where the definition of $\beta$ is used
\begin{equation}
\label{135}
\beta=\frac{c^2\delta m-U(\vec R)}{\varepsilon+m_0c^2}\,. 
\end{equation}
Since the Dirac conjugate $\bar{\psi}=\psi^{*}\gamma^{0}$,
\begin{equation}
\label{131}
\bar{\psi}\psi=\varphi^{*}\chi+\chi^{*}\varphi=|\Phi|^2-|\Theta|^2.
\end{equation}
The electron density is
\begin{equation}
\label{132}
n=|\Phi|^2+|\Theta|^2.
\end{equation}
Below we consider spherically symmetric electron states. All values in such states depend solely on $R$ and therefore $i\vec\sigma$ term in (\ref{130}) disappears. To be specific one can put 
\begin{equation}
\label{134}
\Phi=\frac{1}{\sqrt{2}}{1\choose 1}F. 
\end{equation}

When the deviation $\delta v$ of $v$ from its equilibrium value $\mu c^2$ is small it follows from (\ref{127}) for $\delta m/m_0=\delta v/\mu c^2$ 
\begin{eqnarray}
\label{133}
&&\left(\nabla^2-\frac{2}{R^{2}_{c}}\right)\frac{\delta m}{m_0}\\
&&=\frac{G^2r_c}{2}\left[F^2-\frac{1}{(1+\varepsilon/m_0c^2)^2}\left(\frac{r_c\nabla F}{1+\beta}\right)^2\right],
\nonumber
\end{eqnarray}
where $r_c=\hbar/m_0c\simeq 3.86\times 10^{-11}cm$ is the electron Compton length and $R_c=\hbar/\mu c\sim 10^{-16}cm$ is the Compton length of the Higgs boson. 

The electron density (\ref{132}) now is
\begin{equation}
\label{137}
n=F^2+\frac{1}{(1+\varepsilon/m_0c^2)^2}\left(\frac{r_c\nabla F}{1+\beta}\right)^2.
\end{equation}
The equation for $F$ follows from (\ref{130})
\begin{equation}
\label{138}
-\nabla^2F+\frac{\nabla\beta}{1+\beta}\nabla F=-\frac{1}{r^{2}_{c}}F+\frac{(\varepsilon-U)^2}{\hbar^2c^2}F,
\end{equation}
where a mass variation in the term $1/r^{2}_{c}$ is not important.
\section{SINGULAR SOLUTION}
\label{sing}
Eqs.~(\ref{133}) and (\ref{138}) are valid in the formal absence of fluctuation fields. This corresponds to some scheme of quantum mechanics. Suppose that, in frameworks of this formalism, the 
electron wave function is singular at the point $R=\sqrt{r^2+z^2}=0$. Below we consider the electron in the atomic potential which is approximately  
\begin{equation}
\label{146}
U(R)=-\frac{Ze^2}{\sqrt{R^2+r^{2}_{N}}}
\end{equation}
at distances smaller than the Bohr radius. Here $r_N\sim 10^{-13}cm$ is the nucleus radius. For lead $r_N\simeq 5.5\times 10^{-13}cm$ \cite{MYE}.

At $R\sim r_c$ one can neglect $\nabla\beta$ term and $U$ in the right-hand side of (\ref{138}). In this case the solution of Eq.~(\ref{138}) takes the form \cite{LANDAU2}
\begin{equation}
\label{149a}
F=\frac{C}{R\sqrt{r_c}}\exp\left(-\frac{R}{\hbar c}\sqrt{m^{2}_{0}c^4-\varepsilon^2}\right),
\end{equation}
where $C$ is a dimensionless constant. We suppose $\varepsilon<m_0c^2$. 

Below we take in mind the case of not small $Z$. For example, for lead $Z=82$ and $Ze^2/\hbar c\simeq 0.60$. At $R<r_c$ the left-hand side of Eq.~(\ref{138}), estimated as $\sim F/R^2$, 
dominates the right-hand side of that equation. Therefore $F$ is mainly a solution of (\ref{138}) without the right-hand side
\begin{equation}
\label{250}
\frac{\partial F}{\partial R}=-C\,\frac{1+\beta(R)}{R^2\sqrt{r_c}}\,,\hspace{0.5cm}R<r_c.
\end{equation}

Under the additional condition $R_c<R$ the gradient term in the left-hand side of Eq.~(\ref{133}) is small. But in right-hand sides of Eqs.~(\ref{133}) and (\ref{137}) the gradient terms dominate. 
This results in the mass correction
\begin{equation}
\label{151}
\frac{\delta m(R)}{m_0}=\frac{G^2}{4}r_cR^{2}_{c}n(R),\hspace{0.3cm}R_c<R<r_c\,,
\end{equation}
where the electron density 
\begin{equation}
\label{152}
n(R)=\frac{C^2}{(1+\varepsilon/m_0c^2)^2}\,\frac{r_c}{R^4}\,,\hspace{0.3cm}R_c<R<r_c\,.
\end{equation}

From Eqs.~(\ref{151}) and (\ref{152}) we see how the singularity in the electron distribution is connected with the singularity of the electron mass in the formal absence of fluctuations. At
$R<r_N$ the potential (\ref{146}) is almost a constant and the contribution to $\nabla\beta$ comes from $\delta m$ term in Eq.~(\ref{135}). Therefore there is the singularity source ($\nabla\beta$ 
term) in (\ref{138}) which behaves as inverse power law. This natural singularity source substitutes the artificial $\delta(\vec R)$.

At distances $R$ shorter than $R_c$ the correction $\delta m/m_0$ becomes large and the left-hand side of the equation (\ref{133}), based on the expansion around the equilibrium value $\mu c^2$ 
of $v$, is not correct. In this situation one should use the $v^3$ term in the left-hand side of Eq.~(\ref{127}). One obtains instead of (\ref{151}) $(\delta m/m_0)^3/R^{2}_{c}=G^2r_cn/2$. Since
$G\sim R_c/r_c$ it follows that 
\begin{equation}
\label{152a}
\frac{\delta m}{m_0}\sim\left(\frac{R_c}{R}\right)^{4/3},\hspace{0.5cm}R<R_c.
\end{equation}
With the evaluation (\ref{152a}) Eq.~(\ref{138}) takes the form
\begin{equation}
\label{152b}
-\frac{1}{R^2}\frac{\partial}{\partial R}\left(R^2\frac{\partial F}{\partial R}\right)-\frac{4}{3R}\,\frac{\partial F}{\partial R}=0,\hspace{0.5cm}R<R_c
\end{equation}
and the electron density at $R\lesssim R_c$ remains the same as (\ref{152}), $n\sim r_c/R^4$.

The second term in (\ref{152b}) is the natural singularity source resulting in the form $F\sim 1/R^{7/3}$. Without that term the singularity would be $F\sim 1/R$ corresponding to the artificial
$\delta(\vec R)$ in the right-hand side that does not exist. The number 4/3 in (\ref{152b}) follows from the form (\ref{127}) with the $v^3$ term. In the Standard Model that power is not exactly
known but the general situation with the natural singularity source remains.
\section{SMOOTHING OF THE SINGULARITY}
\label{smoo}
Under the action of electromagnetic fluctuations an electron ``vibrates'' within the certain region of the size $r_L$. The mean displacement amplitude $\langle\vec u\rangle=0$ but the mean squared 
displacement $\langle u^2\rangle=r^{2}_{L}$. In this case the effective potential can be estimated as 
\begin{equation}
\label{146a}
\langle U(|\vec R-\vec u|)\rangle\simeq U(R)+\frac{\langle u^2\rangle}{6} \nabla^2U(R).
\end{equation}
The second term in (\ref{146a}) is a quantum mechanical perturbation resulting in the Lamb shift of energy levels \cite{WEL,MIG,KOL}. This shift can be interpreted in terms of an electron which 
probes various parts of the potential due to ``vibrations``.

When the electron in the well is characterized by the classical frequency $\Omega$,
\begin{equation}
\label{147}
r^{2}_{L}=\langle u^2\rangle=\frac{2r^{2}_{c}}{\pi}\frac{e^2}{\hbar c}\ln\frac{mc^2}{\hbar\Omega}
\end{equation}
(see Appendix). The Lamb radius $r_L$ is determined by photons with frequencies $\omega$ between the limits set by the electron system $\Omega<\omega<mc^2/\hbar$. The calculations of $r_L$ involves 
the regularization of the power law divergence in quantum electrodynamics \cite{LANDAU3}. In atom $\hbar\Omega\sim me^4/\hbar^2$ corresponds to the rydberg energy \cite{MIG}. Therefore
\begin{equation}
\label{147a}
r^{2}_{L}=\langle u^2\rangle=\frac{4r^{2}_{c}}{\pi}\frac{e^2}{\hbar c}\ln\frac{\hbar c}{e^2}\simeq (0.82\times 10^{-11}cm)^2.
\end{equation}
The expressions (\ref{147}) and (\ref{147a}) relate, with the logarithmic accuracy, to the non-relativistic approach.

In the conventional situation of a smooth wave function in a usual potential the length $r_L$ is smaller than the quantum mechanical uncertainty and hence the Lamb shift, related to the second term 
in (\ref{146a}), is relatively small. 
\begin{figure}
\includegraphics[width=5.3cm]{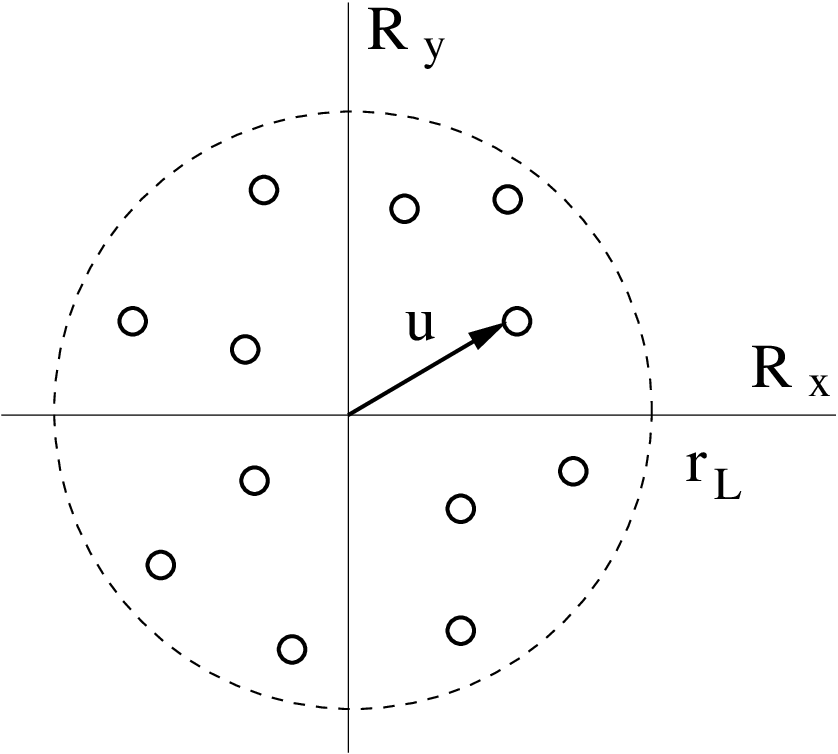}
\caption{\label{fig1}The electron density $n(\vec R-\vec u)$, localized at each fixed $\vec u$ within a circle, is renormalized by fast fluctuations. The subsequent average is due to fluctuations 
of $\vec u$ (various circle positions) which are of lower frequencies.} 
\end{figure}

In our case to obtain the physical electron density from the singular one of Sec.~\ref{sing} one should include fluctuating gauge fields $W^{\pm}_{\mu}$, $Z_{\mu}$, $A_{\mu}$ and the Higgs field 
$h$ which were formally ``switched off'' in Sec.~\ref{sing}. Those fields result in a renormalization, including one of electron mass, and a shift of the singularity position by the fluctuating 
vector $\vec u$. The subsequent average leads to $\langle n(\vec R-\vec u)\rangle$. 

In contrast to quantum electrodynamics, now $\vec u$ depends on all above fluctuating fields. Nevertheless the main contribution to fluctuations of $\vec u$ comes from the massless photon field 
$A_{\mu}$. Massive fields of other gauge bosons and $h$ relate to a shorter fluctuation length. So $\langle u^2\rangle$ can be determined on the basis of quantum electrodynamics. 

At large distances $R$ from the center, $\langle u^2\rangle$ is given by (\ref{147a}). Under the decrease of $R$ this property holds down to $R\sim r_c$ because the Compton radius $r_c$ 
($\simeq 4r_L$) is the border of non-relativistic approach. So at $R>r_c$ the electron density $\langle n(\vec R-\vec u)\rangle$ corresponds to the mean squared displacement 
$\langle u^2\rangle=r^{2}_{L}$ (see Appendix). This distribution at $R>r_c$ looks in the way if singularity centers at $\vec R=\vec u$ would be distributed with the same mean squared displacement. 
That is if $\langle u^2\rangle$ is the same for all $R$ including $R<r_c$. Let us check this property.

At $R>r_c$ the mean squared displacement (\ref{147a}) is formed by fluctuations of the typical time $10^{-15}s$ that is the inverse rydberg energy. Those fluctuations are extremely adiabatic 
compared to processes at $R<r_c$ where the typical time is shorter than $\hbar/mc^2\sim 10^{-21}s$. So the adiabatically varying vector $\vec u$ determines at each moment of time the center
position $\vec R=\vec u$. The subsequent average corresponds to the mean squared displacement (\ref{147a}) of center positions. In other words, the singularity gets smeared out within the sphere 
$R\lesssim r_L$.

This is shown in Fig.~\ref{fig1}, where each instant center position at $\vec R=\vec u$ moves adiabatically compared to the fast dynamics in the vicinity to each center. This fast dynamics extends 
from $mc^2/\hbar$ up to higher frequencies which can be of the Planck scale. So in Fig.~\ref{fig1} the minimal size of each center is not larger than at least the Planck length. 

Through the Fourier component $n_k$ of the function $n(\vec R)$ this average is
\begin{equation}
\label{148a}
\langle n(\vec R-\vec u)\rangle=\int\frac{d^3k}{(2\pi)^3}\,n_k\exp(i\vec k\vec R)\langle\exp(-i\vec k\vec u)\rangle.
\end{equation}
Using the Gaussian average with the condition $\langle u^2\rangle =r^{2}_{T}$, one obtains from (\ref{148a})
\begin{equation}
\label{148b}
\langle n(\vec R-\vec u)\rangle=\int d^{\,3}R_1\,\frac{n(\vec R_1)}{r^{3}_{L}(2\pi)^{3/2}}\exp\left[-\frac{(\vec R-\vec R_1)^2}{2r^{2}_{L}}\right]. 
\end{equation}

As follows from (\ref{148b}) with the normalization condition $\int d^3R_1n(R_1)=1$, the physical electron density, at $R\sim r_L$, is 
\begin{equation}
\label{148c}
\langle n(\vec R-\vec u)\rangle=\frac{1}{r^{3}_{L}(2\pi)^{3/2}}\exp\left(-\frac{R^2}{2r^{2}_{L}}\right). 
\end{equation}

Analogously to Eq.~(\ref{148a}), the physical mass correction $\langle\delta m(\vec R-\vec u)/m_0\rangle$ is expressed through $R^{2}_{1}\delta m(\vec R_1)/m_0$. Since this function has the maximum
at $R_1\sim R_c$, 
\begin{equation}
\label{148d}
\Big\langle\frac{\delta m(\vec R-\vec u)}{m_0}\Big\rangle\sim \left(\frac{R_c}{r_L}\right)^3\exp\left(-\frac{R^2}{2r^{2}_{L}}\right). 
\end{equation}
Here the preexponential coefficient is on the order of $10^{-15}$. So the electron is localized at small region $r_L\sim 10^{-11}cm$ whereas the extra electron mass at that region is negligible. 
\subsection{Anomalous well and anomalous atoms}
\label{smooA}
Since the electron is localized at the region $R<r_L$, its energy, presented in the form
\begin{equation}
\label{149b}
\sqrt{m^2c^4+\frac{\hbar^2 c^2}{r^{2}_{L}}}\simeq \frac{\hbar c}{r_L}\,,
\end{equation}
enhances. Since phenomena at $R<r_L$ are of the electromagnetic origin, the enhancement of the electron energy $\hbar c/r_L\sim 1MeV$ at that region is compensated by the reduction of zero point
photon energy at the same region (anomalous well)
\begin{equation}
\label{149c}
\sum\frac{\hbar\omega}{2}-\left(\sum\frac{\hbar\omega}{2}\right)_0.
\end{equation}
Here the last term relates to absence of the electron. The first term is spatially dependent through the variable density of states. As a result, the energy (\ref{149c}) corresponds to the narrow 
($\sim 10^{-11}cm$) and deep ($\sim 1MeV$) well. Analogous well is formed in the Casimir effect \cite{LANDAU3} of attraction of two atoms when, in contrast, the well is shallow and wide. 

The depth $\hbar c/r_L$ of anomalous well, formed by the reduction of the vacuum energy, is estimated as 
\begin{equation}
\label{9}
{\rm well\,\,depth}\simeq mc^2\sqrt{\frac{\pi\hbar c}{4e^2}\,\frac{1}{\ln(\hbar c/e^2)}}\simeq 2.4MeV
\end{equation}
and cannot be obtained by the perturbation theory on $e^2/\hbar c$. In other words, all orders of the perturbation theory are accounted for. This means that each state of the total system, with 
the particular energy, is exact. Therefore each state has zero width, or equivalently, it lives infinitely long.
\begin{figure}
\includegraphics[width=4.0cm]{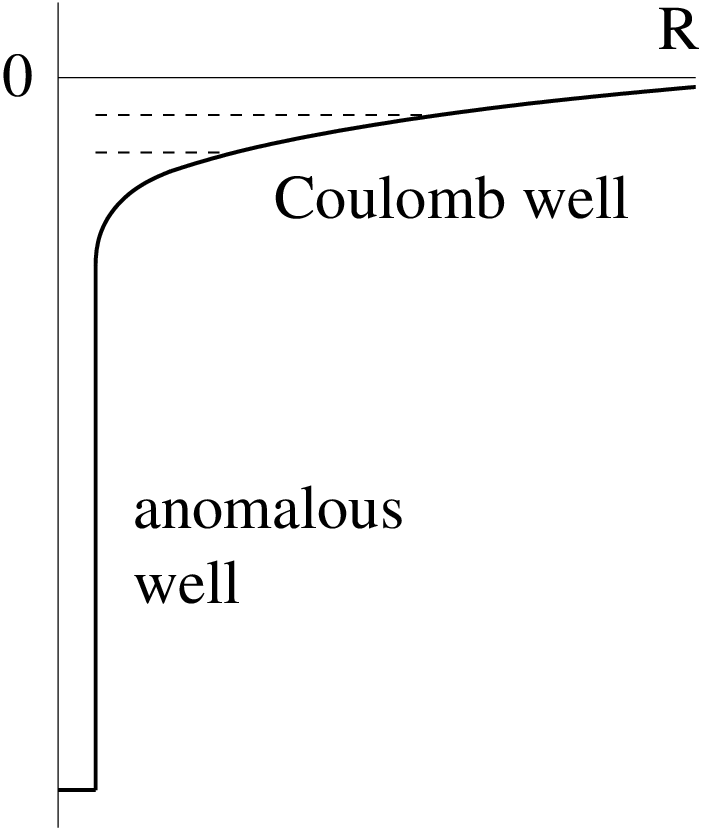}
\caption{\label{fig2}The usual Coulomb well goes over into the narrow ($\sim 10^{-11}cm$) and deep ($\sim 1MeV$) anomalous well. Energy levels in this well are continuous and of zero width. Dashed 
horizontal lines represent energy levels in the initial Coulomb well.} 
\end{figure}

According to quantum mechanics, in a usual potential well energy levels are quantized due to absence of a singularity inside the well. In our case such a condition does not exist since the initial 
singularity, subsequently smeared by fluctuations, relates to a physical state. Therefore instead of levels quantization the spectrum in a well is continuous. There is no contradiction. The 
anomalous electron state has the typical spatial scale $r_L\sim 10^{-11}cm$ corresponding to fast oscillations in space. The typical electron scale in usual atomic systems is $10^3r_L$. So the 
matrix element between those states is of the type $\exp(-1000)$. Hence anomalous states are not formed under usual conditions. 

As follows, the Coulomb potential (\ref{146}) goes over into the anomalous well in Fig.~\ref{fig2} on distances $R<r_L\sim 10^{-11}cm$. The states in the anomalous well are continuous and 
non-decaying (of zero width). The continuous non-decaying spectrum of a particle (attached to an elastic medium) in a potential well is not forbidden in nature. Such spectrum is revealed in 
Ref.~\cite{IVLEV3} on the basis of the exact solution. See also \cite{FEYN1,LEGG,WEISS}. 

The peculiarity of the proposed scheme is that one can omit many details of the mechanism. The conclusion of the singularity is drawn on the basis of mean field approach. The successive 
smearing of the singularity apparently occurs via fluctuating fields with the electromagnetic part as the principal contribution.

So the electron density naturally increases approaching some point where it is not singular. Formation of such anomalous state occurs solely at some restricting macroscopic potential, for example, 
of harmonic type or attractive Coulomb one. For free electrons, as follows from (\ref{147}), $\Omega=0$ and therefore $r_L=\infty$. Hence in this case anomalous state does not exist and there is 
the usual Lehmann representation of the electron propagator according to quantum electrodynamics \cite{LANDAU3}.

It is energetically favorable to capture electrons in the anomalous well getting the energy gain $\sim\hbar c/r_L$ per each. The total energy gain can be approximately estimated as 
\begin{equation}
\label{154a}
\Delta E\simeq -N\left(\frac{\hbar c}{r_L}+\frac{Ze^2}{r_L}\right)+\frac{N^2e^2}{2r_L}\,,
\end{equation}
where $N$ is the number of acquired electrons. The second term is the Coulomb interaction with the nucleus of the charge $Ze$. The third term is due to the Coulomb repulsion of acquired electrons.
The maximal energy gain corresponds to maximal $N$ which cannot be larger than $Z$ because otherwise the confining potential, providing a finite $r_L$, disappears. Putting $N=Z$, one obtains for 
the total binding energy of the anomalous atom
\begin{equation}
\label{154b}
\Delta E\simeq -Z\,\frac{\hbar c}{r_L}\left(1+Z\,\frac{e^2}{2\hbar c}\right).
\end{equation}
For lead $Z=82$ and therefore $\Delta E\simeq -250MeV$. 

The size of the anomalous atom is of $10^{-11}cm$. Energy levels in this atom are continuous and of zero width. So, without external perturbations, the anomalous atom can live infinitely long with 
any energy but not necessary in the ground state after release of the energy ($\ref{154b}$).
\section{CREATION OF ANOMALOUS STATES}
\label{cre} 
When some perturbation has a short spatial scale, comparable with $r_L$, the probability of anomalous state creation is not exponentially small. That perturbation can be a charge density rapidly 
varying in space. This is clarified in the following way. Before the average on $\vec u$ that perturbation quantum mechanically transfers with a not small probability a usual atomic state, with the 
shifted argument $\vec R-\vec u$, to one rapidly varying in space with the same shifted argument. The latter is easily converted into the anomalous one since the both are of the same spatial scale. 
By subsequent fluctuations of $\vec u$ the state turns into the final one. To some extend, this recalls formation of strongly coupled polaron in solids where a potential well is created by phonons 
\cite{KIT}.

When an atom or an ion of the mass $M$ and velocity $V$ is reflected by some potential the interference of incident and reflected de Broglie waves relates to spatially
modulated charge density proportional schematically to $\cos(2MVR/\hbar)$. The typical spatial scale of the charge density is
\begin{equation}
\label{150}
\Delta R=\frac{\hbar}{2MV}\,.
\end{equation}

For example, for the lead atom ($M\simeq 3.44\times 10^{-22}g$) with the velocity $V=1.3\times 10^{5}cm/s$ (speed of sound in lead) the typical scale (\ref{150}) is $1.17\times 10^{-11}cm$. This 
is well compatible with $r_L$. Therefore the perturbation of that spatial scale is effective for creation of anomalous states. 

An interference of incident and reflected de Broglie waves of an atom is also expected in solids. For example, in propagation of a shock wave the lattice site acquires the maximal velocity $V$ at
the shock front position. When the front continues its motion the site returns to the initial position, with the velocity $-V$, due to interaction with other sites. In this process the quantum 
interference of forth and back motions results in the modulation of charge density on the scale (\ref{150}). It is amazing that pure macroscopic processes in condensed matter lead to creation of 
anomalous states where the $MeV$ energy is involved.

The quantum coherence of incident and reflected de Broglie waves has not to not be destroyed by thermal fluctuations. For this reason, the velocity of the macroscopic motion $V$ of lattice sites
should exceed their velocity $V_T=\sqrt{T/M}$ of thermal motion. For lead at room temperature $20^{o}C$ the velocity $V_T$ is 0.08 of the speed of sound. 
\section{MANIFESTATIONS OF ANOMALOUS STATES}
\label{man} 
Since the depth of the anomalous well is on the order of $MeV$ the formation of this well (anomalous atom) occurs during the characteristic time $1/1MeV\sim 10^{-21}s$. That ``jolt'' essentially 
violates electron states. The old stationary states (dashed lines in Fig.~\ref{fig2}) become non-stationary characterized by the flux toward the anomalous well. The distance between the old levels 
is in the $keV$ region. So this process corresponds to the time $1/1keV\sim 10^{-18}s$ and therefore to the quanta emission of the continuous spectrum in the region of $keV$. 

Since the anomalous well depth is $\sim 1MeV$, one can expect an emission of $MeV$ quanta in addition to ones in the $keV$ region. As follows from Sec.~\ref{smooA}, the total binding energy of the 
lead-based anomalous state is around $250MeV$. If, during formation of this state, $(10-20)MeV$ quanta appear one can also expect $\gamma-n$ reactions due to the giant nuclear resonance 
\cite{BAL1,BAL2}. In this case neutron emission can be registered. However the states of an anomalous atom are of zero width and therefore this atom, without external perturbations, is stable with 
any value of electrons energy. So an emission of $MeV$ quanta, in addition to $keV$ ones, requires further studies.

The size of an anomalous atom is three orders of magnitude smaller than conventional one. According to that, there is another unusual aspect of anomalous atoms. If in a part of a solid all atoms 
undergo a transition to the anomalous state that macroscopic region reduces $10^{9}$ times in volume. This process can be qualified as {\it matter collapse}. The collapsed matter (with $10^9$ times 
enhancement of density) looks as a dramatically different concept.

Without external perturbations collapsed matter, as well as an individual anomalous atom, is stable. The energy of this stable state can be in the interval from the initial value (before anomalous 
atoms formation) down to the ground state. One can put a question: what kind of external perturbation triggers off that avalanche releasing $250MeV$ per atom. 

Anomalous threads may exist in vacuum in a magnetic field. In this case the fluctuation radius $r_L$ is also finite if to substitute $\Omega$ in (\ref{147}) by cyclotron frequency. According to
(\ref{147}),
\begin{equation}
\label{155}
r_L\simeq 0.26\sqrt{\ln\frac{4.39\times 10^9}{H(T)}}\times 10^{-11}(cm).
\end{equation}
At $H=1T$ the Lamb radius $r_L\simeq 1.23\times 10^{-11}cm$. The spectrum of those anomalous states is continuous (no transverse quantization) that contrasts to Landau levels in magnetic field.
\begin{figure}
\includegraphics[width=7.0cm]{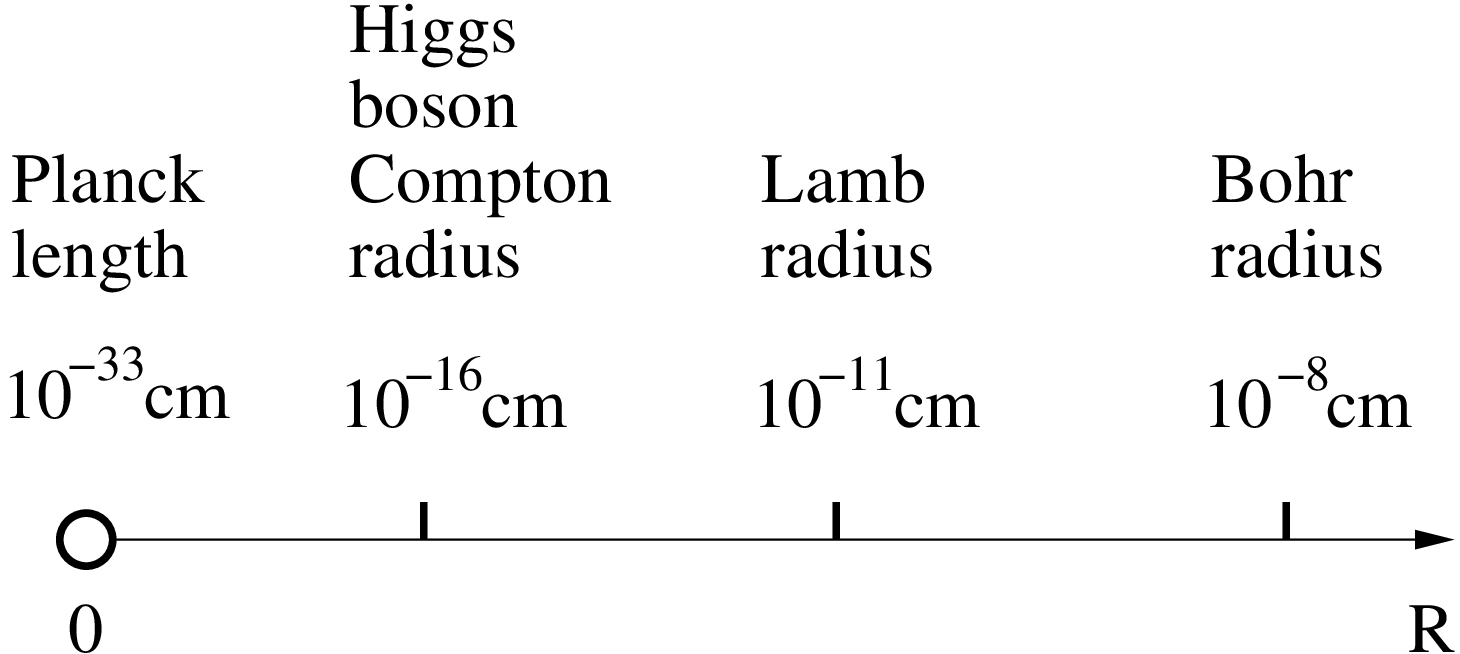}
\caption{\label{fig3} Spatial scales of anomalous electron state. The Planck length is $\sqrt{G_g\hbar/c^3}$, where $G_g$ is the gravitational constant. The Compton radius of the Higgs boson 
$R_c\sim \hbar/\mu c$, where $\mu$ is the mass of the Higgs boson. The Lamb radius $r_L\sim (\hbar/mc)\sqrt{e^2/\hbar c}$. The Bohr radius is $\hbar^2/(me^2)$.}
\end{figure}
\section{DISCUSSIONS}
\label{disc} 
The Schr\"{o}dinger wave equation holds on the scale of the Bohr radius. This equation, at all  $R\neq 0$, can have the formal solution which is $\psi\sim 1/R$ at small $R$. This solution does not 
exist in the whole space since it requires the non-physical singularity source $\delta(\vec R)$. One can try to ``blow up'' the region of small $R$ by involving mechanisms outside the validity of 
Schr\"{o}dinger approach. The scheme of related spatial scales is shown in Fig.~\ref{fig3}. At distances $R$, less than the electron Compton length $r_c$ (between the Lamb radius and the Bohr 
radius), the Dirac wave equation holds but the $\delta$-singularity source remains to be point like. 

The inclusion of the electron-photon interaction, as in quantum electrodynamics, just washes out the $\delta$-source making this non-physical term existing at a finite spatial region of the Lamb
radius $r_L\sim r_c\sqrt{e^2/\hbar c}$. 

Something unusual happens only when we go down to smaller distances, namely, to the Compton length $R_c\sim 10^{-16}cm$ of the Higgs boson. At these distances, in formal absence of fluctuations, 
the singular electron distribution produces the singular part of the electron mass. The latter serves as a natural singularity source (instead of the artificial $\delta(\vec R)$) in the wave 
equation for the bare electron. 

So the singularity of the electron density naturally survives down to the small length which is the border of applicability of the Standard Model. This scenario relates to the bare electron, that
is if to formally remove gauge bosons $W^{\pm}$, $Z$, and $A$, together with the fluctuating part of the Higgs field. With those fields the real state is a superposition of ones with various
singularity positions $\vec R=\vec u$. The true electron density includes the average $\langle n(\vec R-\vec u)\rangle$ with respect to all fluctuating positions $\vec u$. In the usual case this
would correspond to the Lamb effect.

Sweeping of $\vec u$, at a fixed $R$, provides a contribution also from short distances, where the Standard Model is not valid. Nevertheless the physical electron density continues to those 
short distances remaining smooth. The guarantee for that is existence of the fundamental minimal length scale (the Planck length $l$) which serves as cutoff for a possible singularity. The 
subsequent internal scales of anomalous state is shown in Fig.~\ref{fig3}. 

So the electron density naturally increases approaching some point where it is not singular. Formation of such anomalous state occurs solely at some restricting macroscopic potential, for example, 
of harmonic type or attractive Coulomb one. Otherwise the electron is smeared out over the infinite scale and anomalous state does not exist. Anomalous state corresponds to the electron in the thin 
($r_L\sim 10^{-11}cm$) and deep ($c\hbar/r_L\sim 1MeV$) well created by the local reduction of zero point photon energy and localized in the center of a usual atom. 

The energy spectrum in the anomalous well is continuous and non-decaying that is with zero width. The continuous non-decaying spectrum of a particle (attached to an elastic medium) in a potential 
well is not forbidden in nature. Such spectrum is revealed in Ref.~\cite{IVLEV3} on the basis of the exact solution. See also \cite{FEYN1,LEGG,WEISS}.

The possibility in condensed matter of $MeV$ energy states, related to electrons, is surprising. Also it is extremely unusual that those states are formed by macroscopic perturbations which are
shock waves or strong acoustic pulses in a solid. The characteristic time $\Delta t$ of them is approximately the ratio of the width of the shock wave and the velocity of this wave 
\begin{equation}
\label{156}
\Delta t\simeq\frac{10\AA}{10^5cm/s}=10^{-12}s.
\end{equation}
This time is extremely adiabatic compared to the inverse frequency $1/\omega\sim 10^{-18}s$ of emitted $keV$ quanta (X-rays). The usual mechanism of atomic excitation up to $keV$ energies (by the
$\Delta t$ ``jolt'' with a subsequent $\omega$-quanta emission) has the formal probability $\exp(-A)$ where $A\simeq \omega\Delta t\sim 10^{6}$. So $keV$ quanta emission is paradoxical from the 
usual standpoint which deals with a combination of conventional effects. 

Since the states in a anomalous atom are long-living, one can put a question on lasing in such systems. 

The total binding energy of the lead-based anomalous state is around $250MeV$. If, during formation of this state, $(10-20)MeV$ quanta appear one can expect $\gamma-n$ reactions due to the giant 
nuclear resonance. In this case neutron emission can be registered. However the states of an anomalous atom are of zero width and therefore this atom, without external perturbations, is stable 
with any value of electrons energy. So an emission of $MeV$ quanta, in addition to $keV$ ones, requires further studies.
 
The size of an anomalous atom is three orders of magnitude smaller than conventional one. According to that, there is another unusual aspect of anomalous atoms. If in a part of a solid all atoms 
undergo a transition to the anomalous state that macroscopic region reduces $10^{9}$ times in volume. This process can be qualified as {\it matter collapse}. The collapsed matter (with $10^9$ 
times enhancement of density) looks as a dramatically different concept. 

Without external perturbations collapsed matter, as well as an individual anomalous atom, is stable. The energy of this stable state can be in the interval from the initial value (before anomalous 
atoms formation) down to the ground state. One can put a question: what kind of external perturbation triggers off that avalanche releasing $250MeV$ per atom. In this case $36mg$ of lead would
release $4.18\times 10^9J$ originating from a reduction of zero point electromagnetic energy. 

The singular solution of Eq.~(\ref{138}) may be not of the form (\ref{149a}), that has the singularity at the point $R=0$, but of the type $\ln r$ having the singularity on the line $z=0$. The
anomalous state, in the form of thread of $r_L$ radius, can also be formed around that linear singularity. Anomalous threads are likely responsible for the unusually universal features of resistance 
of superconductors \cite{MIL}. 

One can put a question about anomalous states related to quarks. Their mass generation and mixing are also due to the Higgs mechanism with the assistance of Yukawa terms.
\section{CONCLUSIONS}
\label{conc}
Anomalous atoms, of the radius $\sim 10^{-11}cm$ and the energy well of $\sim 1MeV$ depth, are proposed. The wells are due to the spatial variation of zero point electromagnetic energy. These 
anomalous states, from the formal standpoint of quantum mechanics, correspond to a singular solution of a wave equation produced by the non-physical $\delta(\vec R)$ source. The resolution of the 
tiny region around the formal singularity shows that the state is physical. The creation of such state in an atomic system is of the formal probability $\exp(-1000)$. The probability becomes not 
small under a perturbation which rapidly varies in space, on the scale $10^{-11}cm$. In condensed matter such perturbation may relate to acoustic shock waves. In this process the short scale is the 
length of the standing de Broglie wave of a reflected lattice atom. Under electron transitions in the anomalous well (anomalous atom) $keV$ X-rays are expected to be emitted. A macroscopic amount 
of anomalous atoms, of the size $10^{-11}cm$ each, can be formed in a solid resulting in collapsed matter with $10^{9}$ times enhanced density. The collapsed matter (with $10^9$ times enhancement 
of density) looks as a dramatically different concept. 

\acknowledgments
I thank J. L. Diaz Cruz, J. Engelfried, V. P. Gudkov, J. Knight, E. B. Kolomeisky, and M. N. Kunchur for discussions and remarks. This work was supported by CONACYT through grant number~237439.

\appendix*
\section{FLUCTUATION SMEARING OF THE ELECTRON DISTRIBUTION}
Below instead of the usual quantum electrodynamics the model of multi-dimensional quantum mechanics, describing the electron-photon system, is considered. Photons can be treated as an infinite set 
of harmonic oscillators \cite{LANDAU3}. In the method, proposed in Refs.~\cite{FEYN1,LEGG} and developed in further publications (see for example \cite{WEISS}), the Lagrangian of the total system
\begin{eqnarray}
\nonumber
&&L=\frac{m}{2}(\dot{x}^2-\Omega^2x^2)+\frac{\rho}{2L}\sum_k(|\dot{R}_k|^2-\omega^{2}_{k}|R_k|^2)\\
\label{A1}
&&-\frac{x}{L}\sum_kc_kR_k
\end{eqnarray}
depends on ``photon'' coordinates $R_k$, where $R_{-k}=R^{*}_{k}$ and $\omega_k=ck$. The coordinate $x$ refers to the electron. The summation occurs on $-\infty<n<\infty$ with $k=2\pi n/L$ where 
$L$ is the system length. For simplicity we use one dimension as in Refs.~\cite{FEYN1,LEGG} and the harmonic potential $m\Omega^2x^2/2$ for the electron coordinate $x$. The transition to three 
dimensions is easy. The cross-term in (\ref{A1}) describes the ``electron-photon'' interaction. The real coefficients $c_k=c_{-k}$ are specified below. 

The transition from the classical description to the quantum one is clear \cite{LEGG}. One should convert (\ref{A1}) into the Hamiltonian with the substitution of the type 
$m\dot{x}\rightarrow -i\hbar\partial/\partial x$. In quantum electrodynamics one can solve the wave equation with electromagnetic potentials as given functions of space-time and then average on 
them. In our case this is equivalent to solution of the wave equation 
\begin{equation}
\label{A2}
i\hbar\,\frac{\partial\psi}{\partial t}=-\frac{\hbar^2}{2m}\,\frac{\partial^2\psi}{\partial x^2}+\left[\frac{m\Omega^2x^2}{2}-xf(t)\right]\psi,
\end{equation}
where $f(t)=\sum c_kR_k/L$ is a given function of $t$, and the subsequent average on fluctuating variables $R_k$. 

The solution of the Schr\"{o}dinger equation (\ref{A2}) has the form \cite{PER}
\begin{eqnarray}
\nonumber
&&\psi(x,t)=\varphi(x-u,t)\exp\bigg[\frac{im}{\hbar}(x-u)\dot{u}\\
\label{A3}
&&+i\int^t\frac{dt_1}{\hbar}\left(\frac{m}{2}\dot{u}^2-\frac{m\Omega^2x^2}{2}+xf\right)\bigg],
\end{eqnarray}
where the function $\varphi(x,t)$ obeys the equation without $f$
\begin{equation}
\label{A4}
i\hbar\,\frac{\partial\varphi}{\partial t}=-\frac{\hbar^2}{2m}\,\frac{\partial^2\varphi}{\partial x^2}+\frac{m\Omega^2x^2}{2}\varphi. 
\end{equation}
The shift $u(t)$ satisfies the classical equation of motion
\begin{equation}
\label{A5}
m\ddot{u}+m\Omega^2u=f(t).
\end{equation}
In our case $f$ depends linearly on fluctuating variables $R_k$ and $\langle u\rangle=0$. Photon degrees of freedom $R_k$ satisfy the classical equations
\begin{equation}
\label{A6}
\rho(\ddot{R}_k+\omega^{2}_{k}R_k)=-c_ku.
\end{equation}
One can easily substitute into (\ref{A5}) the expression of $R_k$ through $u$ from (\ref{A6}). The result is
\begin{eqnarray}
\label{A7}
&&m\ddot{u}(t)+m\Omega^2u\\
\nonumber
&&+\frac{2}{\pi}\int^{t}_{-\infty}dt_1\dot{u}(t_1)\int^{\infty}_{0}d\omega\eta(\omega)\cos\omega(t_1-t)=0,
\end{eqnarray}
where the summation rule and the viscosity coefficient are
\begin{equation}
\label{A8}
\sum_k=\frac{L}{\pi c}\int^{\infty}_{0}d\omega,\hspace{0.5cm}\eta(\omega_k)=\frac{c^2(\omega_k)}{2\rho c\omega^{2}_{k}}\,.
\end{equation}
We use the notation $c(\omega_k)=c_k$.

As follows from (\ref{A3}), the instant electron density $n[x-u(t)]=|\varphi(x-u)|^2$ is determined by the shifted solution of the Schr\"{o}dinger equation without the electron-photon coupling. This
coupling enters the game through the shift $u(t)$ of that solution. The function $\varphi(x)$ can be either an eigenfunction or one with a tendency to be singular at $x\rightarrow 0$. 

Smearing in space of the physical density $\langle n(x-u)\rangle$, after averaging on $u$, is determined by $\langle u^2\rangle$. This parameter turns to zero if to formally put $\eta(\omega)=0$ 
since $u$ corresponds to electron ``vibrations'' due to its interaction with photon environment. 

It is most easier to calculate $\langle u^2\rangle$ by fluctuation-dissipation theorem 
\begin{equation}
\label{A9}
\langle u^2\rangle=\frac{i\hbar}{2\pi}\int^{\infty}_{-\infty}\cot\frac{\hbar\omega}{2T}\,\frac{d\omega}{m\omega^2-m\Omega^2+i\eta(\omega)\omega}
\end{equation}
by subtracting the part which remains if to formally put $\eta=0$. 

Let us choose
\begin{equation}
\label{A10}
\eta(\omega)=\frac{2e^2}{3c^3}\,\omega^2.
\end{equation}
Then the classical equation (\ref{A7}) turns to the three-dots-equation of the classical field theory
\begin{equation}
\label{A11}
m\ddot{u}-\frac{2e^2}{3c^3}\dddot{u}+m\Omega^2u=0,
\end{equation}
which is well discussed in textbooks, see for example \cite{LANDAU1}.

As follows from (\ref{A9}) at $T=0$, 
\begin{equation}
\label{A12}
\langle u^2\rangle=\frac{\hbar}{2m\omega}\left[1-\frac{1}{\pi m}\int^{\infty}_{0}\frac{\eta(\omega)d\omega}{(\Omega+\omega)^2}\right].
\end{equation}
With the expression (\ref{A10}) the integral in (\ref{A12}) is divergent and, according to rules of quantum electrodynamics, it should be regularized by subtraction of the divergent part 
\cite{LANDAU3}
\begin{equation}
\label{A13}
\frac{1}{(\Omega+\omega)^2}\rightarrow\frac{1}{(\Omega+\omega)^2}-\frac{1}{\omega^2}\,. 
\end{equation}
Doing this regularization in (\ref{A12}) and subtracting the $\eta=0$ part, one obtains
\begin{equation}
\label{A14}
\langle u^2\rangle=\frac{\hbar}{\pi m^2}\int^{\infty}_{0}\frac{\eta(\omega)d\omega}{\omega(\Omega+\omega)^2}
\end{equation}
with $\eta(\omega)$ from (\ref{A10}). The integral in (\ref{A14}) diverges only logarithmically. This divergence is not required a further regularization since it is related to the non-relativistic
approach used. Due to that the integration in (\ref{A14}) is restricted by $mc^2/\hbar$ \cite{MIG}. Under this condition, multiplying (\ref{A14}) by 3 due to the dimensionality, we obtain the
expression (\ref{147}).

The shift of the ground state energy of hydrogen atom (the Lamb shift), caused by the second term in Eq.~(\ref{146a}) \cite{WEL,MIG,KOL}, is 
\begin{equation}
\label{A15}
\Delta E=\frac{2me^4}{3\hbar^3}\frac{\langle u^2\rangle}{r^{2}_{B}}\,,
\end{equation}
where $r_B=\hbar^2/(me^2)$ is the Bohr radius. The comparison of (\ref{A15}) with the exact expression for $\Delta E$, following from quantum electrodynamics \cite{LANDAU3}, shows that the 
expressions (\ref{147a}) is exact within the logarithmic accuracy. For calculations beyond that accuracy the non-relativistic approach used is not sufficient.

\end{document}